\documentclass[prb,twocolumn,showpacs,showkeys]{revtex4}

\usepackage{graphicx}
\usepackage{subfigure}
\usepackage{amssymb}
\usepackage[dvips]{color}

\begin{document}
\title{Electronic structure and crystal phase stability of palladium hydrides}

\author{Abdesalem Houari}
\email[Corresponding author:]{abdeslam.houari@univ-bejaia.dz}
\affiliation{Theoretical Physics Laboratory, 
             Department of Physics, 
             University of Bejaia, 
             Bejaia, Algeria}

\author{Samir F.\ Matar}
\email[]{matar@icmcb-bordeaux.cnrs.fr}
\affiliation{CNRS, ICMCB, Universit\'e de Bordeaux,
             33600 Pessac, France}

\author{Volker Eyert}
\email[]{veyert@materialsdesign.com}
\affiliation{Materials Design SARL, 92120 Montrouge, France}

\date{\today}

\begin{abstract}
The results of electronic structure calculations for a variety of 
palladium hydrides are presented. The calculations are based on 
density functional theory and used different local and semilocal 
approximations. The thermodynamic stability of all structures as 
well as the electronic and chemical bonding properties are addressed. 
For the monohydride, taking into account the zero-point energy is 
important to identify the octahedral Pd-H arrangement with its larger 
voids and, hence, softer hydrogen vibrational modes as favorable 
over the tetrahedral arrangement as found in the zincblende and 
wurtzite structures. Stabilization of the rocksalt structure is due 
to strong bonding of the $ 4d $ and $ 1s $ orbitals, which form a 
characteristic split-off band separated from the main $ d $-band 
group. Increased filling of the formerly pure $ d $ states of the 
metal causes strong reduction of the density of states at the Fermi 
energy, which undermines possible long-range ferromagnetic order 
otherwise favored by strong magnetovolume effects. For the dihydride, 
octahedral Pd-H arrangement as realized e.g.\ in the pyrite structure 
turns out to be unstable against tetrahedral arrangemnt as found in 
the fluorite structure. Yet, from both heat of formation and 
chemical bonding considerations the dihydride turns out to be less 
favorable than the monohydride. Finally, the vacancy ordered defect 
phase $ {\rm Pd_3H_4} $ follows the general trend of favoring the 
octahedral arrangement of the rocksalt structure for Pd:H ratios 
less or equal to one. 
\end{abstract}


\maketitle

\section {Introduction}
\label{sec:intro}

Metal hydrides have attracted a lot of attention in the last decades 
due to a variety of exciting applications especially with regard to 
the environmentally friendly storage and use of energy.  In this 
context, they have proven as very promising candidates for hydrogen 
storage, fuel cells applications \cite{rosi03,jain10}, and heat storage 
for the solar-energy industry \cite{felderhoff09}. These auspicious 
perspectives are complemented by strong interest from the point of view 
of fundamental research inter alia because of their role as possible 
candidates for high-temperature superconductivity \cite{ashcroft04}. 
Since hydrogen insertion can act as negative pressure it may also cause 
a switching between different electronic or magnetic states, as has been 
found, e.g., in intermetallic cerium compounds \cite{reviewCe}.

Especially the palladium-hydrogen system has been investigated since 
long \cite{graham1869,lewis60} due to the fact that palladium allows 
to absorb comparatively large amounts of hydrogen 
\cite{lewis67,muller68,westlake78,alefeld78,lewis82,flanagan91,fukai93,flanagan94,manchester94}. 
For this reason, much interest has centered about the maximum amount of 
hydrogen to be absorbed in the metal as well as the possible hydrogen 
positions in the parent structure. This has initiated several structural 
characterizations as well as theoretical studies. Many investigations 
focused on the question, which open voids in the underlying face-centered 
cubic lattice of palladium metal are most likely to be occupied by hydrogen. 
Especially, for the monohydride the exact position of the hydrogen atoms 
was a matter of long dispute. According to neutron diffraction data, 
hydrogen atoms occupy the octahedral interstices of the face-centered 
cubic lattice of Pd leading to the rocksalt structure 
\cite{worsham57,flanagan91,fukai93,manchester94}. However, depending on 
temperature a transfer of substantial amounts of hydrogen to the 
tetrahedral interstices, leading to the zincblende structure, has been 
inferred from neutron diffraction data \cite{ferguson65}. For the 
palladium-deuterium system a partial occupation of the tetrahedral site 
was also reported and attributed to the lower zero-point energy of the 
D interstitial as compared to H \cite{pitt03}. 

Due to their increased hydrogen content as compared to the monohydrides, 
transition metal dihydrides have also come into focus quite early, see, 
{\em e.g.}\ the work by Switendick and references therein 
\cite{switendick70,switendick71}. In this case, there was agreement on 
the fluorite structure. More recently, the formation of superabundant 
vacancy structures was observed at high temperatures and high hydrogen 
pressures, possibly the most prominent example being $ {\rm Pd_3H_4} $, 
which arises from the monohydride on removing Pd and forms an ordered 
structure \cite{fukai94,degtyareva09,harada07}. This vacancy formation 
is accompanied by a compression of the lattice. 

Due to the high interest in the palladium-hydrogen system, a number 
of theoretical studies have been performed ranging from simplified 
non-self-consistent calculations using atomic potentials 
\cite{switendick70,switendick71,gelatt78,gupta78} to fully self-consistent 
relativistic state-of-the-art electronic structure calculations as 
based on density functional theory and (semi-)local approximations 
to the exchange-correlation functional 
\cite{wkg,methfessel82,chan83,elsaesser91,caputo03,dyer06,isaeva11}. 
Nevertheless, even the early work by Switendick revealed considerable 
distortions of the Pd band structure on hydrogenation and allowed to 
rule out previous rigid-band considerations \cite{switendick70,switendick71}. 
In particular, based on his finding of only a partial charge transfer 
from hydrogen to the metal, Switendick was the first to abandon the 
then popular anion and proton models, which were based on an 
ionic picture. These findings were lateron confirmed by more refined 
calculations \cite{gelatt78,gupta78,wkg,chan83}. More recent theoretical 
work centered about the issue of the hydrogen site in PdH, which was 
motivated by the afore mentioned disputes \cite{caputo03,dyer06,isaeva11}. 

As is well known from a number of studies, hydrogen insertion in metals 
may induce long-range magnetic order via strong magnetovolume effects. 
This is especially relevant for paramagnetic palladium, which is at the 
verge of a ferromagnetic instability.   
Yet, in general there is a strong competition between the 
magnetovolume effects enhancing magnetization and the effects of hydrogen 
bonding to the underlying metal substructure leading to the suppression 
of the magnetic order \cite{matar03,matar10}. 

The present work employed electronic structure calculations as based 
on density functional theory and focusses on the structure-property 
relationships of several palladium hydrides. We thus follow two main 
goals: First, emphasis is on the thermodynamic stability of a variety 
of palladium hydrides with perfect stoichiometry and different crystal 
structures. This includes three possible structures of the monohydride 
PdH: the observed rocksalt structure, the zincblende structure, and the 
hexagonal wurtzite structure. In addition, the hydrogen over-stoichiometric 
palladium dihydride $ {\rm PdH_2} $ is considered in both the fluorite 
and the related pyrite structure. Finally, the above mentioned recent 
experiments suggested to investigate the new ordered Pd-defect phase 
$ {\rm Pd_3H_4} $. The second goal of this work centers about the 
calculation of the electronic and magnetic properties as well as 
chemical bonding indicators, which lead to a deeper understanding of 
the various phases.

\section {Computational Methods}       
\label{sec:comp}

The first principles calculations were based on density functional 
theory \cite{hohenberg64,kohn65} with exchange-correlation effects 
accounted for by both the local and semilocal approximation. Two 
complementary tools were employed: 

In a first step, the Vienna ab-initio simulation program (VASP) as 
implemented in the MedeA$ \textsuperscript{\textregistered} $ 
computational environment of Materials Design was used to perform 
total-energy and force calculations aiming at an optimization of 
the structures \cite{vasp,medea}. The exchange-correlation functional 
was considered at three different levels, namely, within the local 
density approximation \cite{perdew81}, the generalized gradient 
approximation (GGA) by Perdew, Burke, and Ernzerhof (PBE) 
\cite{perdew96a}, and the PBEsol scheme recently devised especially 
for the improved treatment of solids \cite{perdew08}. The 
single-particle equations were solved using the projector-augmented
wave (PAW) method \cite{paw,vasppaw} with a plane-wave basis with a
cutoff of 326.203 eV. The Brillouin zone was sampled using a 
Monkhorst-Pack mesh with a spacing below 0.1 \AA$^{-1}$ leading to 
$ 27 \times 27 \times 27 $ {\bf k}-points for the rocksalt and 
zincblende structures as well as $ 26 \times 26 \times 13 $ 
{\bf k}-points for the wurtzite structure \cite{monkhorst76}. 
Furthermore, $ 25 \times 25 \times 25 $ and $ 16 \times 16 \times 16 $ 
{\bf k}-points were used for the dihydride and $ {\rm Pd_3H_4} $, 
respectively. Phonon spectra were calculated using the Phonon software 
\cite{parlinski97}, which is likewise part of the 
MedeA$ \textsuperscript{\textregistered} $ computational environment 
and used the forces evaluated with the VASP code. 

Once the optimized structures were known, calculations of the electronic 
structure and the chemical bonding properties were carried out using 
the full-potential augmented spherical wave (ASW) method in its 
scalar-relativistic implementation \cite{aswrev,aswbook}. Exchange and 
correlation effects were considered using the GGA parametrization of 
Wu and Cohen \cite{wu06}. 
In the ASW method, the wave function is expanded in atom-centered
augmented spherical waves, which are Hankel functions and numerical
solutions of Schr\"odinger's equation, respectively, outside and inside
the so-called augmentation spheres. In order to optimize the basis set 
and enhance the variational freedom,
additional augmented spherical waves were placed at carefully selected
interstitial sites. The choice of these sites as well as the augmentation
radii were automatically determined using the sphere-geometry optimization
algorithm \cite{SGO}. Self-consistency was achieved by a highly efficient
algorithm for convergence acceleration \cite{mixpap} until the variation 
of the atomic charges was smaller than $ 10^{-8} $ electrons and the 
variation of the total energy was smaller than $ 10^{-8} $ Ryd. Brillouin 
zone integrations were performed using the linear tetrahedron method 
\cite{bloechl94} with 
$ 27 \times 27 \times 27 $ {\bf k}-points for face-centered cubic Pd, 
$ 26 \times 26 \times 26 $ {\bf k}-points for the rocksalt structure, 
$ 25 \times 25 \times 25 $ {\bf k}-points for the zincblende structure, 
and $ 24 \times 24 \times 12 $ {\bf k}-points for the wurtzite structure. 
In addition, $ 24 \times 24 \times 24 $ and $ 15 \times 15 \times 15 $ 
{\bf k}-points were used for the dihydride and $ {\rm Pd_3H_4} $, 
respectively. 

In the present work, a new full-potential version of the ASW method was 
employed \cite{aswbook}. In this version, the electron density and related 
quantities are given by spherical harmonics expansions inside the muffin-tin 
spheres while, in the remaining interstitial region, a representation in 
terms of atom-centered Hankel functions is used \cite{methfessel88}. 
However, in contrast to previous related implementations, no so-called 
multiple-$\kappa$ basis set is needed, rendering the method computationally 
nearly as efficient as the original ASW scheme.

\section {Results}
\label{sec:results}

\subsection {Monohydride PdH}
\label{sec:pdh}

As mentioned above, three different structures were considered for the 
monohydride, namely, the rocksalt structure with hydrogen placed at the 
octahedral sites of the face-centered cubic lattice formed by elemental 
Pd, the zincblende structure with hydrogen being at the tetrahedral 
sites, and the hexagonal wurtzite structure, which also has the hydrogen 
atoms at the center of the tetrahedral voids of the Pd sublattice. All 
structures were fully optimized, i.e.\ the lattice vectors as well as 
the internal coordinates were relaxed using VASP. The results are 
summarized in Table \ref{tab:table1}, 
\begin{table}[htb]
\caption{\label{tab:table1}
Calculated equilibrium properties of PdH in the rocksalt (RS), 
zincblende (ZB), and wurtzite (WZ) structures: total energies $ E $ 
per formula unit, lattice constants $ a $ and $ c $, as well as  
hydrogen parameters $ u $ within LDA/PBEsol/PBE. Calculated 
properties of palladium metal and experimental data (from Refs.\ 
\onlinecite{kittel}, \onlinecite{schirber75}, and \onlinecite{fukai93}) 
as well as previous calculational results are added for comparison.} 
\begin{ruledtabular}
\begin{tabular}{l|l|c@{ }c@{ }c|c@{ }c@{ }c}
   &         & \multicolumn{3}{c}{$ E - E_{RS} $ (meV) }
             & \multicolumn{3}{c}{$ a $ (\AA) [, $ c $, $ u $]} \\
\hline
Pd & this work                       &  & &  & 3.84 & 3.87 & 3.94 \\
   & Ref.\ \onlinecite{methfessel82} &  & &  & 3.87 &      &      \\
   & Ref.\ \onlinecite{elsaesser91}  &  & &  & 3.88 &      &      \\
   & Ref.\ \onlinecite{caputo03}     &  & &  &      &      & 3.97 \\
   & Ref.\ \onlinecite{isaeva11}     &  & &  & 3.88 &      &      \\
   & Ref.\ \onlinecite{kittel}       &  & &  & \multicolumn{3}{c}{3.89} \\
\hline
RS & this work                       & \multicolumn{3}{c|}{0}     
                                             & 4.03 & 4.06 & 4.12 \\
   & Ref.\ \onlinecite{methfessel82} &  & &  & 4.12 &      &      \\
   & Ref.\ \onlinecite{switendick87} &  & &  & 4.08 &      &      \\
   & Ref.\ \onlinecite{elsaesser91}  &  & &  & 4.07 &      &      \\
   & Ref.\ \onlinecite{caputo03}     &  & &  &      &      & 4.15 \\
   & Ref.\ \onlinecite{isaeva11}     &  & &  & 4.06 &      &      \\
   & Ref.\ \onlinecite{schirber75}   &  & &  & \multicolumn{3}{c}{4.09} \\
   & Ref.\ \onlinecite{fukai93}      &  & &  & \multicolumn{3}{c}{4.07} \\
\hline
ZB & this work                       & -17 & -25 & -83 
                                             & 4.14 & 4.17 & 4.23 \\
   & Ref.\ \onlinecite{elsaesser91}  &  & &  & 4.17 &      &      \\
   & Ref.\ \onlinecite{caputo03}     &  & &  &      &      & 4.26 \\
\hline
WZ & this work                       & +17 &  +9 & -49
                                             & 2.88 & 2.90 & 2.95 \\
   &                                 &  & &  & 4.93 & 4.97 & 5.02 \\ 
   &                                 &  & &  & 0.36 & 0.36 & 0.36 \\
\end{tabular}
\end{ruledtabular}
\end{table}
which gives calculated lattice parameters and internal parameters (in case 
of the wurtzite structure) arising from the three exchange-correlation 
functionals mentioned above together with available theoretical and 
experimental data from the literature. In all cases, the agreement with 
previous structural data is very good. This is true also for elemental 
Pd, which is considered in the upper part of the table for comparison. 
Especially noteworthy are the experimental results of 
Refs.\ \onlinecite{kittel}, \onlinecite{schirber75}, and 
\onlinecite{fukai93}, which reveal very good agreement with the 
calculations using the PBEsol functional. 

Table \ref{tab:table1} also lists the calculated total energies, which 
are given relative to those of the rocksalt structure, again for all 
three exchange-correlation functionals. Obviously, the zincblende 
structure is found to be most stable for all three functionals. Yet, 
this is in contrast to most experimental findings 
\cite{worsham57,flanagan91,fukai93,manchester94} despite some hints at 
an occupation of the tetrahedral site \cite{ferguson65,pitt03}. 
In addition, it is at variance with the situation of hydrogen in nickel, 
where calculations reveal the octahedral site as the most stable one 
\cite{wimmer08}. 

In order to resolve the issue, we have complemented the calculations 
underlying Table \ref{tab:table1} by calculations of the phonon spectra 
of all three structures, thereby focusing on the use of the PBEsol 
exchange-correlation functional, as this gives the best structural data 
in comparison with experiment. From these calculations, zero-point energies 
of 94, 219, and 214 meV (per formula unit) were obtained for the rocksalt, 
zincblende, and wurtzite structure, respectively. Obviously, addition of 
the zero-point energies reverses the order of total energies found for 
the three different structures, leaving the rocksalt structure as the 
most stable one. Specifically, with the zero-point energies included, 
the zincblende and wurtzite structures have energies, which are 100 and 
129 meV, respectively, above that of the rocksalt structure. 

\begin{figure}[htb]
\includegraphics[width=\columnwidth]{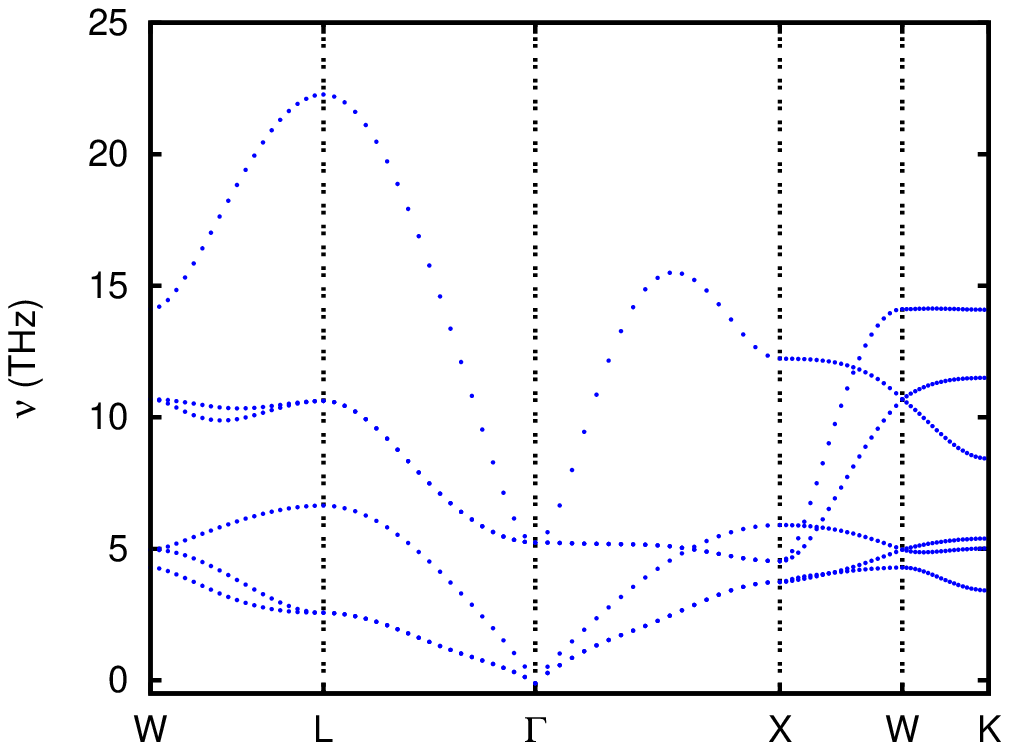}
\includegraphics[width=\columnwidth]{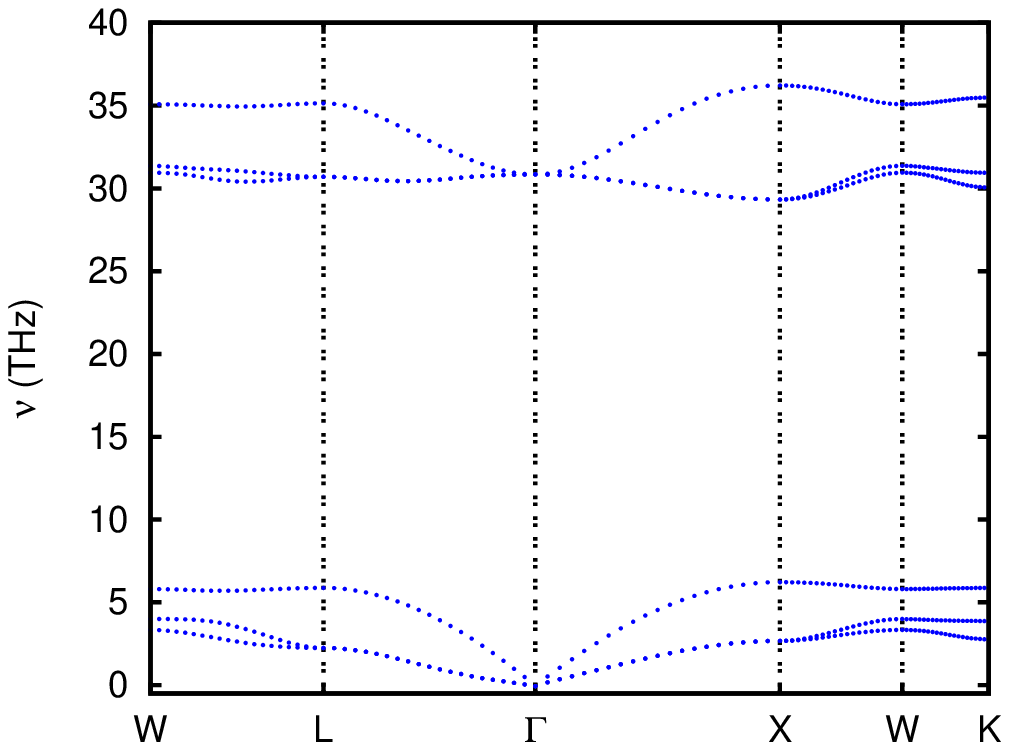}
\caption{Phonon dispersions of PdH in the rocksalt (H in octahedral site, top) 
         and zincblende (H in tetrahedral site, bottom) structure.}
\label{fig:phonPdH}
\end{figure}

The above striking differences can be understood from a closer inspection 
of the phonon dispersions calculated for both the rocksalt and the zincblende 
structure as displayed in Fig.\ \ref{fig:phonPdH}. 
Results for the wurtzite structure are very similar to those obtained 
for the zincblende structure. For the cubic structures the phonon 
dispersions fall into two groups with three phonon branches each. While 
the lower set of branches extends up to about 7\,THz for both structures, 
strong differences are found for the upper group. Specifically, these 
branches are found at frequencies around 32\,THz for the zincblende 
structure, whereas for the rocksalt structure they extend from about 
5\,THz to a maximum of about 22\,THz, thus explaining the much higher 
zero-point energy of the former structure. 

As a deeper analysis reveals, the lower group of branches corresponds to 
phonon modes with a cooperative motion of Pd and H, whereas the higher lying 
branches exclusively involve motion of the rather rigid H and Pd sublattices 
relative to each other and, hence, strongly affect the Pd-H bonds.  In 
this context it is interesting to relate the differences in zero-point 
energy to the differences in Pd-H bond lengths in all three structures. 
We find 2.03 \AA, 1.80 \AA and 1.79/1.81 \AA, respectively, for the 
rocksalt, zincblende, and wurtzite structure (with the first value given 
for the latter structure referring to the bond parallel to the hexagonal 
$ c $ axis). These bond length are much larger than the sum of the covalent 
radii (1.28 \AA and 0.32 \AA) due to the fact that the lattice spacing is 
governed by the Pd-Pd bonds. As a consequence, hydrogen vibrations relative 
to the neighbouring Pd atoms affect the metal-hydrogen bonds to a much 
lesser degree in the rocksalt structure as compared to the zincblende 
structure and are thus found at lower energies. This explains the strong 
difference in zero-point energies, which are roughly inversely proportional 
to the difference between the bond lengths and sum of the covalent radii. 

\begin{figure}[htb]
\includegraphics[width=\columnwidth]{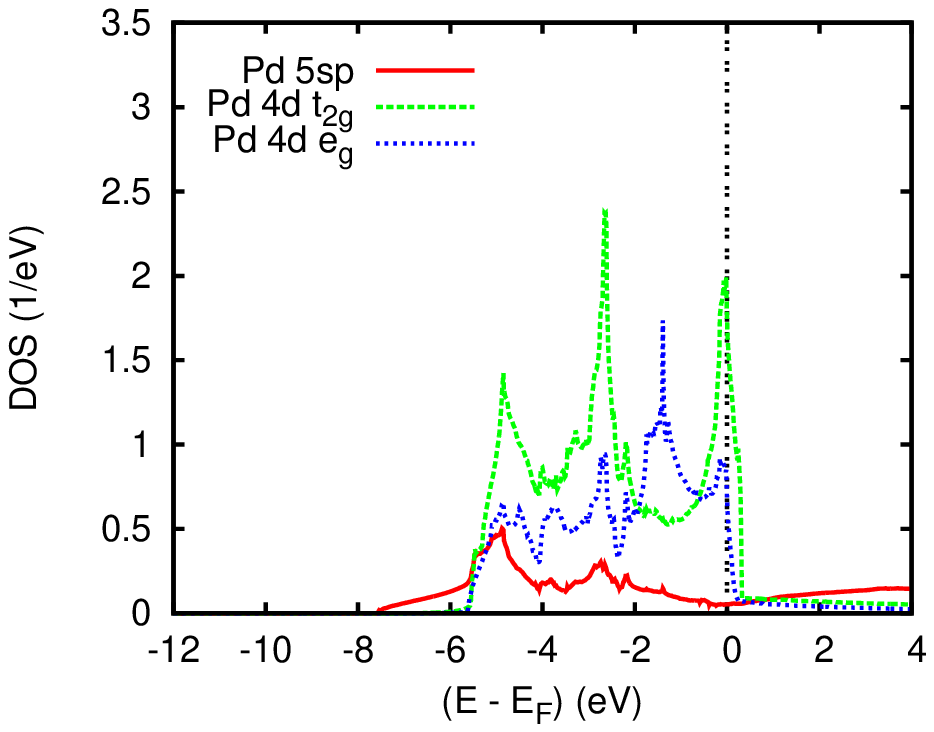}
\includegraphics[width=\columnwidth]{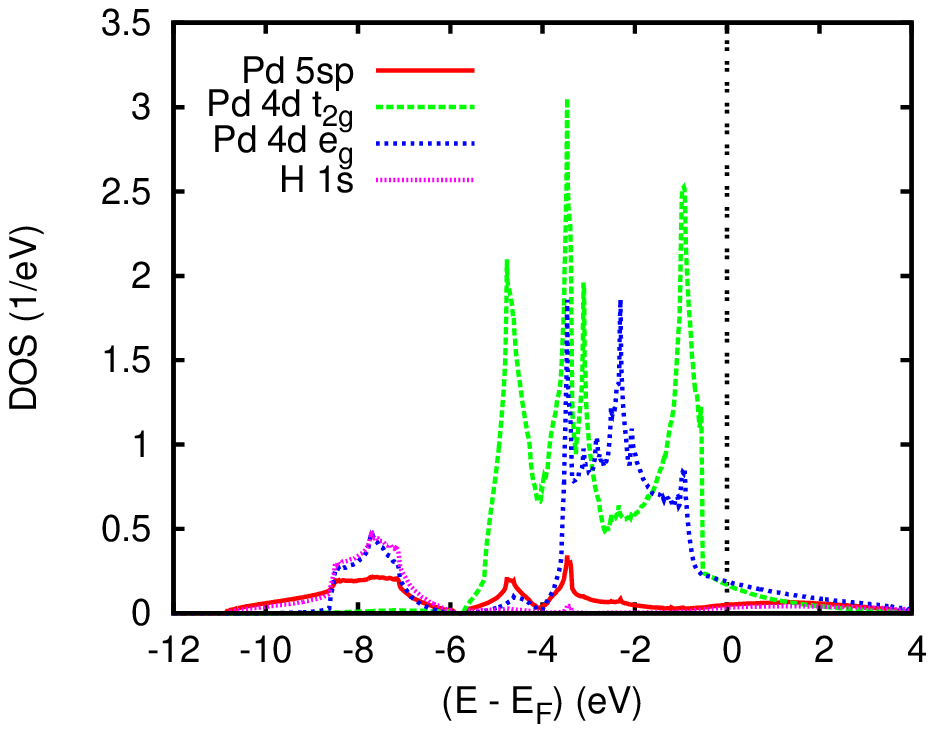}
\includegraphics[width=\columnwidth]{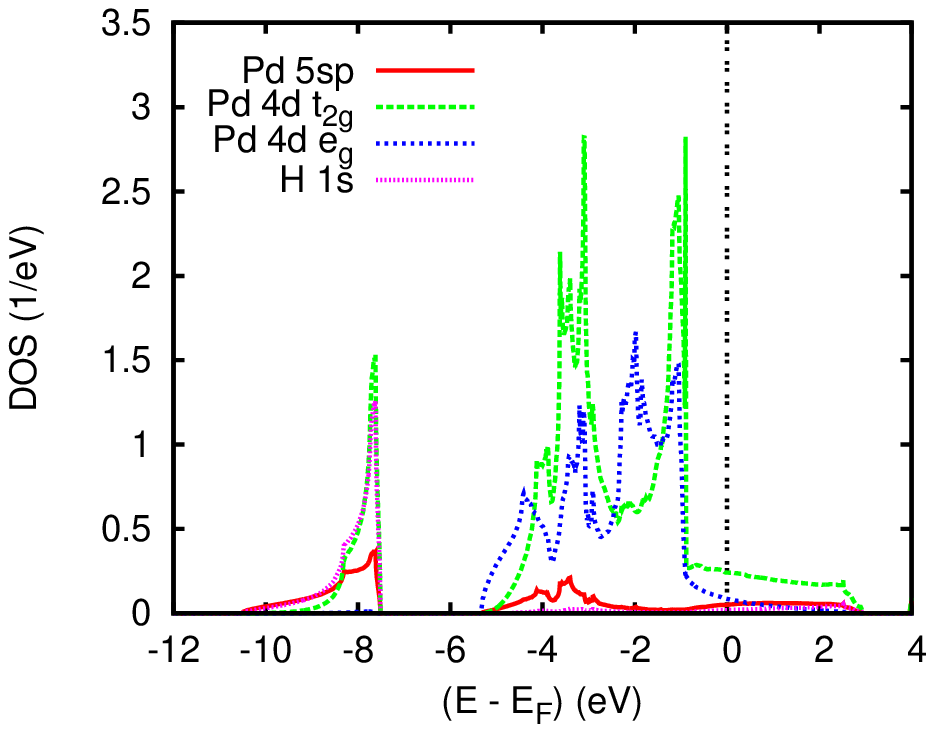}
\caption{Partial densities of states of Pd (top) and PdH in the rocksalt 
         (H in octahedral site, middle) and zincblende (H in tetrahedral 
         site, bottom) structure.}
\label{fig:dosPdPdH}
\end{figure}

In order to discuss the previous results from the electronic structure, 
calculations of the (partial) densities of states were performed 
using the ASW method. The results for elemental Pd as well as PdH 
in the rocksalt and zincblende structures are shown in Fig.\ 
\ref{fig:dosPdPdH}. 
For all three systems we find dominant contributions from the Pd $ 4d $ 
states, which are complemented by small admixtures from the $ 5sp $ 
states especially in the low energy part of the spectrum. 
Two findings are especially noteworthy: First, the main body of the 
Pd $ 4d $ states is much narrower in the hydride structures as compared 
to elemental Pd. Second, for the hydrides a split-off band is observed 
at about $ - 8 $\,eV in very good agreement with photoemission data 
\cite{schlapbach82,bennett82,sinha86}. This band arises from almost equal 
contributions of the Pd $ 4d $ and H $ 1s $ states, which are almost 
identical in shape. The latter fact is a signature of a strong overlap 
of these states. Indeed, as Fig.\ \ref{fig:dosPdPdH} reveals for the 
rocksalt structure, the split-off band comprises only 
those $ 4d $ states, which form strong $ \sigma $-type bonds with the 
$ 1s $ states. In the octahedral arrangement of the rocksalt structure 
these are the $ e_g $ states. In contrast, in the zincblende and wurtzite 
structures the palladium atoms are tetrahedrally coordinated and, hence, 
the strong $ \sigma $-type $ 4d $-$ 1s $ overlap is mediated by the 
$ t_{2g} $ orbitals. This leads to the characteristic shape of the 
split-off band, which is similar for the zincblende and wurtzite 
structures but differs from that obtained for the rocksalt structure. 
Nevertheless, the strong $ \sigma $-type overlap with the almost 
identical shape of the Pd $ 4d $ and H $ 1s $ contributions impressively 
confirms the predominantly covalent metal-hydrogen bonding of palladium 
hydride as put forward already by Switendick \cite{switendick70,switendick71}. 
It explains the fact that the Pd-H phases are rather metastable inasmuch 
as they readily decompose, contrary to very stable ionic hydrides, whose 
archetype is $ {\rm MgH_2} $, where the electron transfer is much larger 
\cite{matar13}. Interestingly, in cubic $ \beta $-$ {\rm MgPd_3H_{0.7}} $ 
low-lying Pd-H bonding states are also found. Yet, there are no noticable 
contributions from Mg in this energy region, which fact reveals the strong 
Pd-preference of hydrogen as well as the tendency to form strong covalent 
Pd-H bonds \cite{kohlmann09}. 

The results obtained for rocksalt structure PdH are very similar to 
those shown by previous authors \cite{gupta78,wkg,methfessel82,chan83}. 
In particular, Williams, K\"ubler, and Gelatt attribute the split-off 
band to the attractive potential arising from the interstitial hydrogen 
proton, which leads to strong overlap of the $ 4d $ and $ 1s $ states 
and thus lays ground for the chemical bonding of the hydrides. They 
find a very similar behavior as well for NiH and the neighboring 
hydrides in the $ 3d $ and $ 4d $ series. 

\begin{figure}[htb]
\includegraphics[width=\columnwidth]{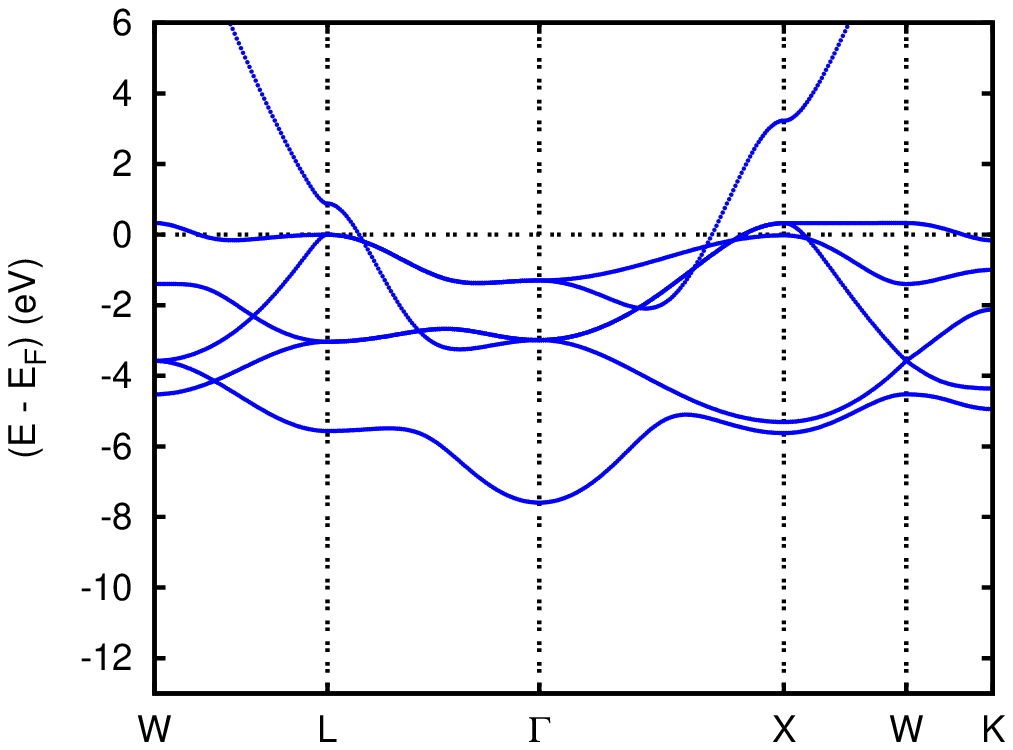}
\includegraphics[width=\columnwidth]{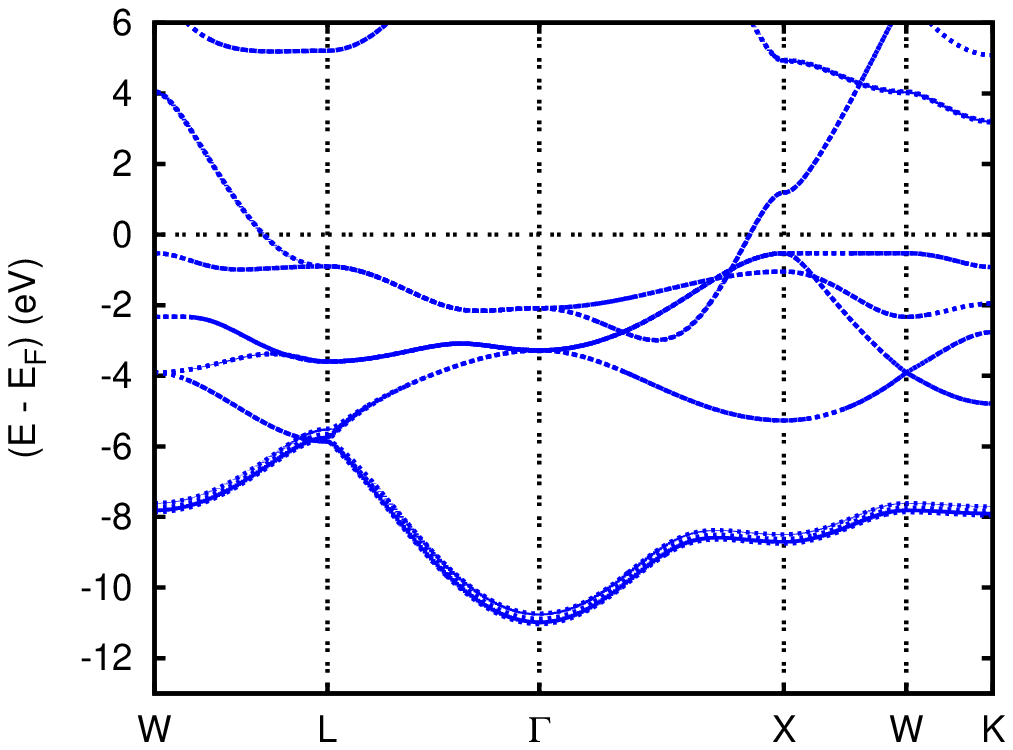}
\caption{Electronic bands of Pd (top) and PdH in the rocksalt (H in 
         octahedral site, bottom) structure. The width of the bars 
         given with each band is a measure of the H $ 1s $ contribution 
         to the respective wave function.}
\label{fig:bndPdPdH}
\end{figure}

It is interesting to compare the band structures of elemental Pd and PdH 
in the rocksalt structure, which are displayed in Fig.\ \ref{fig:bndPdPdH}. 
They clearly reveal the downshift of the lowest band of Pd metal by about 
2-3\,eV in good agreement with all previous calculations. However, while 
the early non-self-consistent calculations find this band completely 
separated from 
the higher lying bands \cite{switendick70,switendick71,gupta78}, we find 
a band crossing close to the L point in our calculation in accordance 
with previous self-consistent calculations \cite{chan83}. The low-lying 
band at L seems to disperse up to $ - 4 $\,eV at the W point. However, the 
dominant H $ 1s $ character, which is indicated in Fig.\ \ref{fig:bndPdPdH}, 
always remains with the lowest band. 

The downshift of the $ d $ band group is most obvious from the (almost) 
dispersionless highest $ d $ bands along the lines W-L and X-W-K, which 
straddle the Fermi energy in Pd but are found completely occupied in PdH. 
As a result, as has been also mentioned by Gupta and Freeman 
\cite{gupta78}, the Fermi surface becomes much clearer and assumes the 
shape known from the nobel metals. Another consequence of this lowering 
is the drastic reduction of the density of states at the Fermi level 
from 2.47 states/eV to 0.46 states/eV in perfect agreement with previous 
calculations \cite{switendick70,switendick71,chan83}. 

\begin{figure}[htb]
\includegraphics[width=\columnwidth]{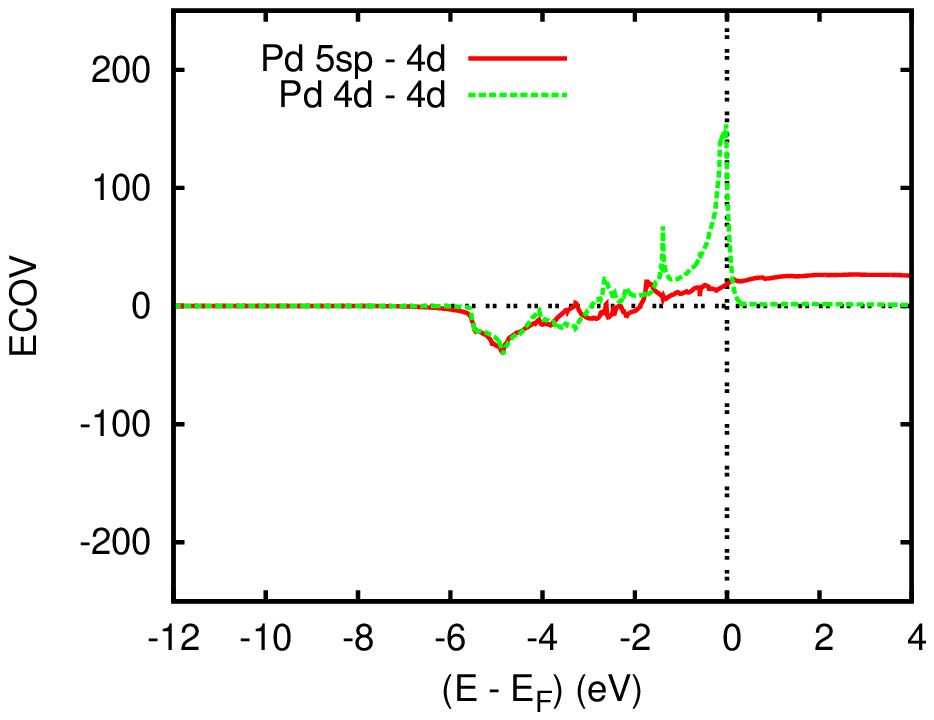}
\includegraphics[width=\columnwidth]{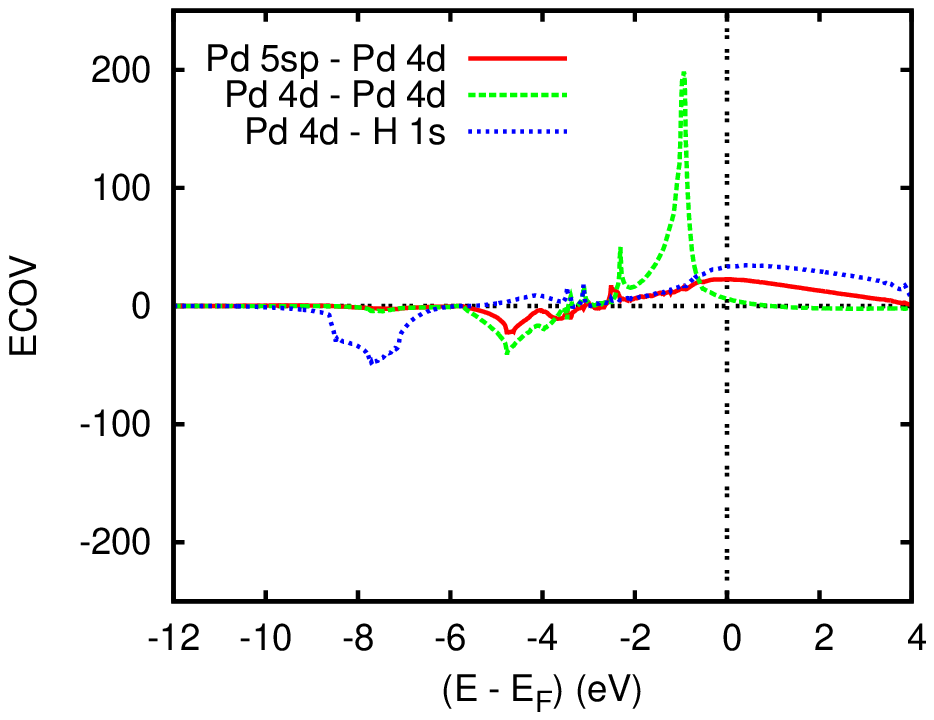}
\includegraphics[width=\columnwidth]{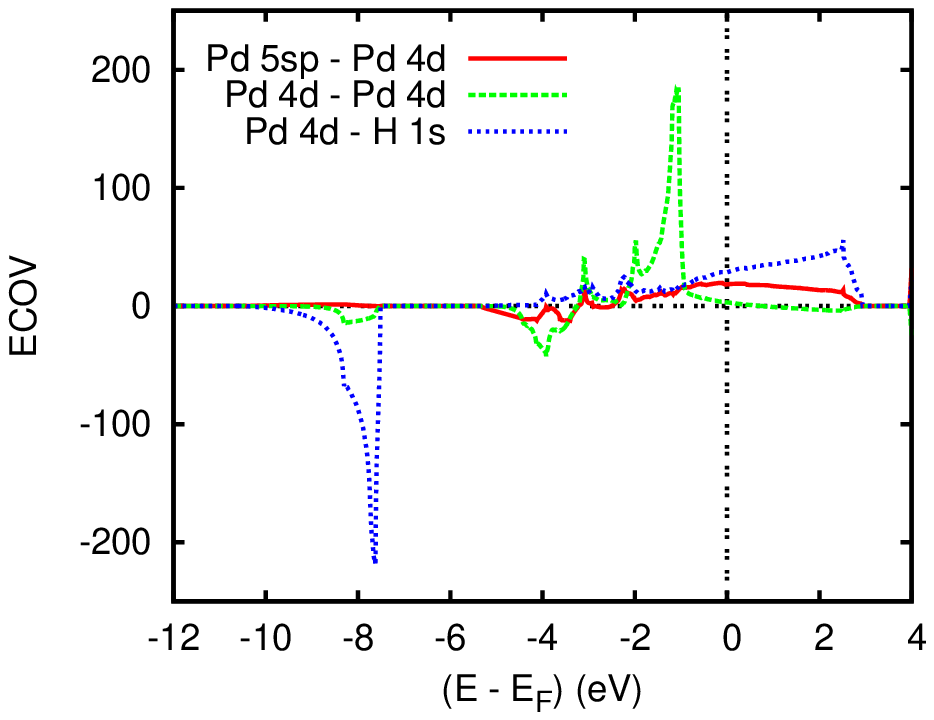}
\caption{Partial covalence energies of Pd (top) and PdH in the rocksalt 
         (H in octahedral site, middle) and zincblende (H in tetrahedral 
         site, bottom) structure.}
\label{fig:coopPdPdH}
\end{figure}

We complement the above considerations with a discussion of the chemical 
bonding properties as they are provided by the crystal orbital overlap 
population (COOP) or the more recent covalence energy scheme (ECOV) used 
here \cite{hoffmann88,boernsen99}. Calculated results for the latter are 
displayed in Fig.\ \ref{fig:coopPdPdH}. 
Note that negative and positive contributions to the covalence energy 
point to bonding and antibonding orbitals, respectively. Note also 
that by definition the covalence energy is a dimensionless quantity. 
For elemental Pd, $ d $-$ d $ bonding and antibonding contributions 
found in the lower and upper range of the $ 4d $ band, respectively, 
nearly outweigh each other. In contrast, $ sp $-$ d $ bonding contributions 
extend up to about $ - 2 $\,eV and thus lead to a stabilization of the 
metal. These essential findings are recognized also in the hydride 
structures. However, due to the upshift of the Fermi level, the 
respective antibonding states attain a higher occupation and would 
destabilize the metallic framework were it not for the hydrogen bonding. 
While the overlap of the Pd $ 5sp $ states with the hydrogen orbitals 
plays only a negligible role, the $ 4d $-$ 1s $ overlap as expected 
turns out to be essential for the stability of the hydride. Obviously, 
the split-off band comprises the bonding contributions, whereas the 
antibonding parts set in only at much higher energies and start to 
counterbalance the bonding states just above the Fermi energy. 

\begin{figure}[htb]
\includegraphics[width=\columnwidth]{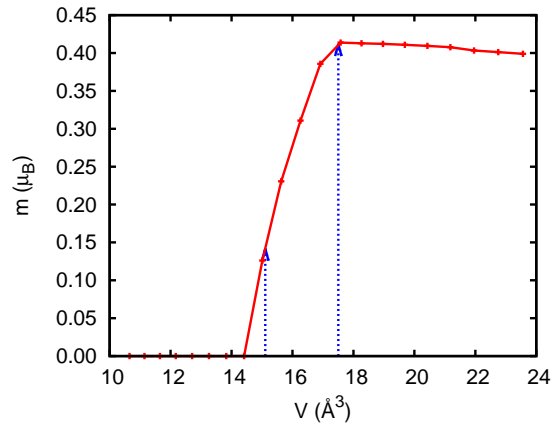}
\caption{Magnetic moment of Pd per unit cell as a function of volume. 
         Connecting lines serve as guide to the eye only; arrows 
         at 15.1\,$ {\rm \AA^3} $ and 17.5\,$ {\rm \AA^3} $ mark the 
         equilibrium volumes of Pd and rocksalt PdH, respectively.}
\label{fig:momPd}
\end{figure}

\begin{figure}[htb]
\includegraphics[width=\columnwidth]{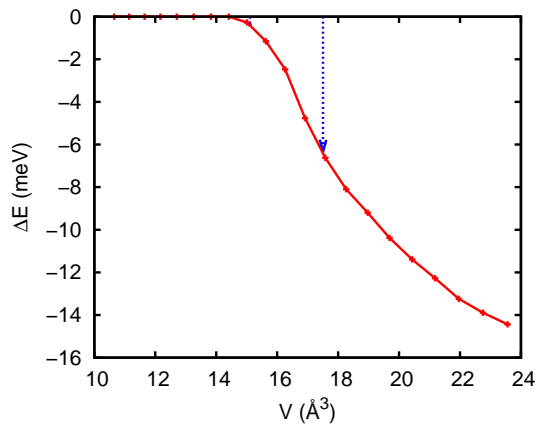}
\caption{Total energy difference between spin-polarized and spin-degenerate 
         calculations of Pd as a function of volume. Connecting lines serve 
         as guide to the eye only; arrows at 15.1\,$ {\rm \AA^3} $ and 
         17.5\,$ {\rm \AA^3} $ mark the equilibrium volumes of Pd and 
         rocksalt PdH, respectively.}
\label{fig:ediffPd}
\end{figure}

As well known, transition-metal hydrides may show a tendency towards 
long-range magnetic order due to their expanded volumes 
\cite{matar03,matar10}. In particular, previous calculations for elemental 
Pd were interpreted in terms of a possible ferromagnetic instability at 
expanded lattice constants \cite{castro91,hong08}. For this reason, we 
performed additional spin-polarized calculations in order to check for 
stable ferromagnetic order. The results of spin-polarized ASW calculations 
are displayed in Fig.\ \ref{fig:momPd}.  
The corresponding lowering of the total energy on including spin-polarization 
is captured in Fig.\ \ref{fig:ediffPd}.  
According to these findings, Pd metal is indeed on the verge of 
developing a finite magnetization; at the equilibrium lattice 
constant, it amounts to $ \approx 0.14 \mu_B $. On expanding 
the lattice, the magnetization sharply increases until it reaches its 
saturation value of about $ 0.4 \mu_B $ at the lattice constant of PdH. 
These findings may be easily understood from the strong peak in the 
density of states at the Fermi energy as seen in Fig.\ \ref{fig:dosPdPdH}, 
which according to Stoner theory may cause a ferromagnetic instability. 
On expanding the lattice this peak narrows and becomes stronger, which 
fact is in favor of the energy gain due to magnetic ordering as deduced 
from the Stoner picture. Hence, were it only for the volume expansion 
effects brought by hydrogen insertion, PdH would develop stable 
ferromagnetic order. However, as discussed in connection with the 
densities of states shown in Fig.\ \ref{fig:dosPdPdH}, hydrogen insertion, 
in addition to expanding the lattice, pushes the Fermi energy to the 
energy range above the strong $ 4d $ peaks where the density of states 
is small. As a consequence, the magnetic instability is suppressed. From 
a different point of view, this suppression of magnetic ordering may 
be assigned to the quenching of the local magnetic moments by the spin 
pairing coming with the Pd-H bonding. Given the densities of states of 
Pd metal shown in Fig.\ \ref{fig:dosPdPdH} this result is very likely 
to hold even at a reduced hydrogen content. Nevertheless, it is important to 
note that the energy gain on turning on spin-polarization and expanding 
the lattice is of the order of 10\,meV and, hence, any long-range ordering 
would occur at very low temperatures only.

\subsection {Dihydride $ {\bf PdH_2} $}
\label{sec:pdh2}

Following the early work by Switendick, we extended the present study 
to the dihydride $ {\rm PdH_2} $, which he assumed to crystallize in 
the cubic fluorite structure coming with a tetrahedral coordination 
of the Pd atoms by hydrogen. In the present structure, we alternatively  
considered the pyrite structure as a starting point for the structural 
relaxations. This was motivated by its close relationship to the fluorite 
structure as well as by a variety of exciting results recently obtained 
for pyrite-type materials \cite{eyert98,weihrich03a,weihrich03b,houari10}. 
Yet, in contrast to the fluorite structure, in the pyrite structure the 
metal atoms are octahedrally coordinated and the hydrogen atoms form 
closely coupled dimers parallel to the $ \langle111\rangle $ direction. 
To be specific, the pyrite structure is based on a simple cubic lattice 
with space group {\em Pa$\bar{3}$} ($ T_h^6 $) (see Ref.\ 
\onlinecite{eyert98} and references therein). The atoms are located at 
the Wyckoff positions (4a) and (8c), which are special cases of the 
general position (24d): $ {\rm \pm (x,y,z)} $,
$ {\rm \pm (\frac{1}{2}-x,-y,\frac{1}{2}+z)} $,
$ {\rm \pm (-x,\frac{1}{2}+y,\frac{1}{2}-z)} $,
$ {\rm \pm (\frac{1}{2}+x,\frac{1}{2}-y,-z)} $,
(and cyclic permutations of x, y, and z) with $ {\rm x = y = z = 0} $ 
and $ {\rm x = y = z \approx 0.4} $, respectively. The fluorite structure 
arises when the (8c) position is given by $ {\rm x = y = z = 0.25} $, 
in which special case the crystal symmetry is raised to the full cubic 
group {\em Fm$\bar{3}$m} ($ O_h^5 $). As for the monohydride structures, 
full optimizations including the lattice parameter and the hydrogen 
position were performed using VASP. However, the pyrite structe proved 
unstable inasmuch as optimization of this structure eventually led to 
the fluorite structure. This is most probably due to strong H-H bonds 
coupling the characteristic hydrogen dimers in the pyrite structure, 
which undermine the Pd-H bonding and thus lead to a positive energy of 
formation. 

As a consequence, only the fluorite structure was included in the 
subsequent analysis of the electronic structure. The lattice parameter 
of the optimized structure was 4.37, 4.39, and 4.45 {\AA} as obtained 
from LDA, PBEsol, and PBE calculations, respectively. In addition, the 
heat of formation was calculated as $ -0.54 $\,eV, $ -0.22 $\,eV, and 
$ +0.10 $\,eV, respectively, all per formula unit. These values should 
be compared to the heats of formation of PdH calculated from these three 
schemes, which amount to $ -0.51 $\,eV, $ -0.36 $\,eV, and $ -0.16 $\,eV, 
respectively (the LDA value for PdH compares nicely with the value of 
$ -0.53 $\,eV reported by Isaeva {\em et al.}\cite{isaeva11}). Hence, 
within the GGA schemes, PdH is more stable than $ {\rm PdH_2} $, which 
is even on the verge of forming a thermodynamically stable compound. 

\begin{figure}[htb]
\includegraphics[width=\columnwidth]{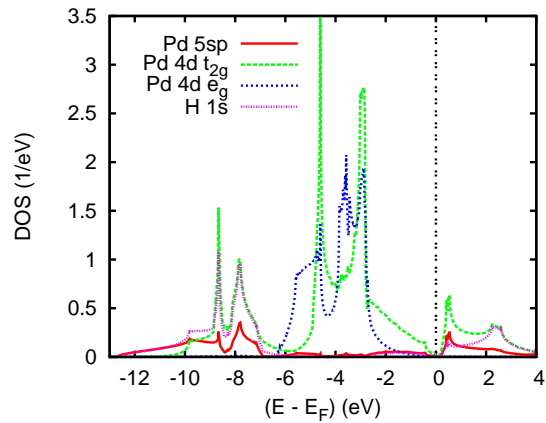}
\caption{Partial densities of states of $ {\rm PdH_2} $ in the fluorite  
         structure.}
\label{fig:dosPdH2}
\end{figure}

The calculated (partial) densities of states arising from the subsequent  
ASW calculations are shown in Fig.\ \ref{fig:dosPdH2}. 
As for the monohydrides, we find dominant contributions from the 
Pd $ 4d $ states, which are complemented by small admixtures from 
the $ 5sp $ states. In addition, considerable contributions from the 
H $ 1s $ states are observed, especially between $ -12 $ and $ -7 $\,eV 
and above the Fermi energy. Contributions from the Pd $ 4d $ states 
in these energy ranges are of similar size as the H $ 1s $ contributions 
and of almost identical shape, indicating the strong covalent bonding 
of these states. According to Fig.\ \ref{fig:dosPdH2} and very similar 
to the situation in zincblende-type PdH this $ \sigma $-type bonding 
is maintained exclusively by the $ t_{2g} $ manifold, whereas the 
$ e_g $ orbitals have no overlap with the H $ 1s $ states at all. 
Two additional points are especially noteworthy: In contrast to the 
partial densities of states of zincblende-type PdH as shown in the 
lower part of Fig.\ \ref{fig:dosPdPdH} the split-off band of $ {\rm PdH_2} $ 
between $ -12 $ and $ -7 $\,eV falls into two main peaks, which both 
carry a considerable weight of the H $ 1s $ orbitals. This is due to 
the higher amount of electrons contributed from the hydrogen atoms. 
Yet, as is obvious from the similar $ 4d $ and $ 1s $ weights in that 
energy range, these electrons are not simply transferred to the metal 
atoms as in an ionic scenario but rather participate in the covalent 
bonding. 

Second, as a detailed comparison of the partial densities of states of 
the dihydride to those of zincblende-type PdH reveals, the two groups of 
bands ranging from $ -12 $ to $ -7 $\,eV and from $ -7 $\,eV to the 
Fermi energy correspond to the two groups of bands displayed in the 
lower part of Fig.\ \ref{fig:dosPdPdH} for zincblende PdH. Hence, in 
$ {\rm PdH_2} $, we observe an additional third group of bands, which 
are found above $ {\rm E_F} $. These bands may be regarded as the 
antibonding Pd $ 4d $-H $ 1s $ states, which together with their bonding 
partners between $ -12 $ and $ -7 $\,eV embrace the central group of 
metal $ 4d $ states. 

\begin{figure}[htb]
\includegraphics[width=\columnwidth]{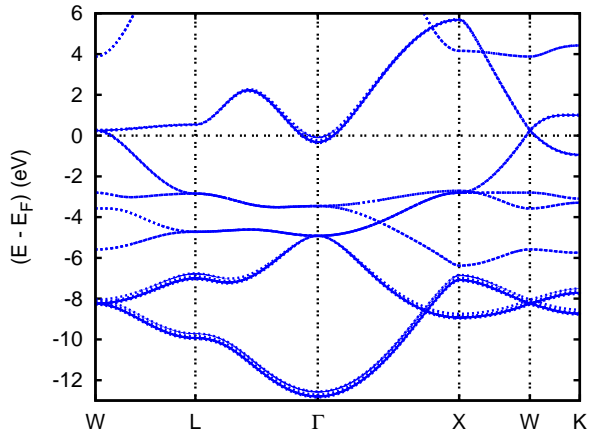}
\caption{Electronic bands of $ {\rm PdH_2} $. Contributions of the 
         H $ 1s $ orbitals are highlighted. The width of the bars 
         given with each band is a measure of the H $ 1s $ contribution 
         to the respective wave function.}
\label{fig:bndPdH2}
\end{figure}

The gross picture is complemented by a closer inspection of the band 
structure as given in Fig.\ \ref{fig:bndPdH2}, 
which reveals the stronger distortion of the bands of elemental Pd as 
compared to PdH due to the larger hydrogen content of the dihydride. 
In particular, as mentioned above, now the two lowest bands of the 
Pd $ 4d $ complex comprise a finite contribution of the H $ 1s $ 
orbitals. In addition, the second band shows strong hybridization 
with a new single band, which starts just below the Fermi energy and 
extends up to about 6\,eV with the hybridization being strongest at 
the $ \Gamma $ point. This new band leads to the above mentioned 
additional contribution to the density of states. 

\begin{figure}[htb]
\includegraphics[width=\columnwidth]{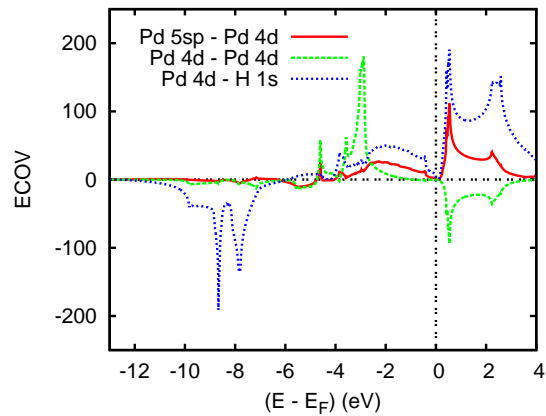}
\caption{Partial covalence energies of $ {\rm PdH_2} $ in the fluorite 
         structure.}
\label{fig:coopPdH2}
\end{figure}

As before, we complement the previous analysis of the electronic structure 
and partial densities of states with an investigation of the chemical 
bonding properties in terms of the partial covalence energies, which are 
displayed in Fig.\ \ref{fig:coopPdH2}. 
Overall, the results are quite comparable to those of zincblende-type 
PdH. Again, the Pd $ 4d $-H $ 1s $ overlap plays an essential role 
for stabilizing the dihydride, with the split-off bands between 
$ -12 $ and $ -7 $\,eV comprising the bonding contributions. However, 
due to the downshift of the central group of $ 4d $ bands, occupation 
of the corresponding antibonding states sets in at a lower energy. As 
a consequence, we observe the destabilization of dihiydride as compared 
to PdH. Finally, occupied antibonding $ 5sp $-$ 4d $ and $ 4d $-$ 4d $ 
orbitals even enhance this effect. In passing, we mention the somewhat 
unexpected negative $ d $-$ d $ covalence energies above the Fermi 
energy. However, it should be taken into account that in this energy 
region there is a strong similarity of the Pd $ 4d $ and $ 5sp $ partial 
densities of states as revealed by Fig.\ \ref{fig:dosPdH2}, pointing to 
interatomic covalent $ 4d $-$ 5sp $ bonding, which according to to Fig.\ 
\ref{fig:coopPdH2} adds to the $ d $-$ d $ interactions and counterbalances 
these.

\subsection{$ {\rm \bf Pd_3H_4} $ phase}
\label{sec:pd3h4}

As noted in the introduction, recent experiments have shown that palladium 
hydride tends to form palladium vacancies at about 700-800\,$ ^{\circ} $C 
and 5\,GPa hydrogen pressure. This induces a phase transition to a stable 
$ {\rm Pd_3H_4} $ composition, which forms an ordered structure arising 
from PdH on removal of e.g.\ the Pd atoms at the corners of the cubic cell  
\cite{harada07,isaeva11}. Following Isaeva {\em et al.}\ we investigated 
$ {\rm Pd_3H_4} $ in both the so modified rocksalt and zincblende structures  
\cite{isaeva11}. However, while these authors focused throughout on the 
thermodynamic stability of the two structures, we here concentrate on 
the electronic structure and the structure-property relationships. 

Yet, again the investigation was initiated by structure optimizations 
employing the VASP code. The results as obtained from different 
approximations to the exchange-correlation functional are summarized 
in Table \ref{tab:table3}.
\begin{table}[htb]
\caption{\label{tab:table3}
Calculated equilibrium properties of $ {\rm Pd_3H_4} $ in the rocksalt 
(RS) and zincblende (ZB) structures: total energies $ E $ per formula 
unit and lattice constants $ a $ within LDA/PBEsol/PBE. Experimental 
data (from Ref.\ \onlinecite{fukai94}) are added for comparison.} 
\begin{ruledtabular}
\begin{tabular}{l|l|c@{ }c@{ }c|c@{ }c@{ }c}
   &         & \multicolumn{3}{c}{$ E - E_{RS} $ (meV) }
             & \multicolumn{3}{c}{$ a $ (\AA)} \\
\hline
RS & this work                       & \multicolumn{3}{c|}{0}     
                                             & 3.88 & 3.91 & 3.96 \\
   & Ref.\ \onlinecite{isaeva11}     &  & &  & 3.92 &      &      \\
   & Ref.\ \onlinecite{fukai94}      &  & &  & \multicolumn{3}{c}{4.02} \\
\hline
ZB & this work                       & 447 & 406 & 254 
                                             & 3.95 & 3.98 & 4.06 \\
   & Ref.\ \onlinecite{isaeva11}     &  & &  & 4.04 &      &      \\
\end{tabular}
\end{ruledtabular}
\end{table}
As before, we find an increase of the lattice constant on going from LDA 
to PBEsol and PBE. Our results are in reasonable agreement with those 
of Isaeva {\em et al.}\ but deviate from the larger experimental value. 
However, they do reproduce the volume reduction on going from PdH to 
the vacancy phase. Finally, we find the zincblende-derived structure at 
much higher energies than the rocksalt-derived structure, again in 
agreement with experimental data. 

\begin{figure}[htb]
\includegraphics[width=\columnwidth]{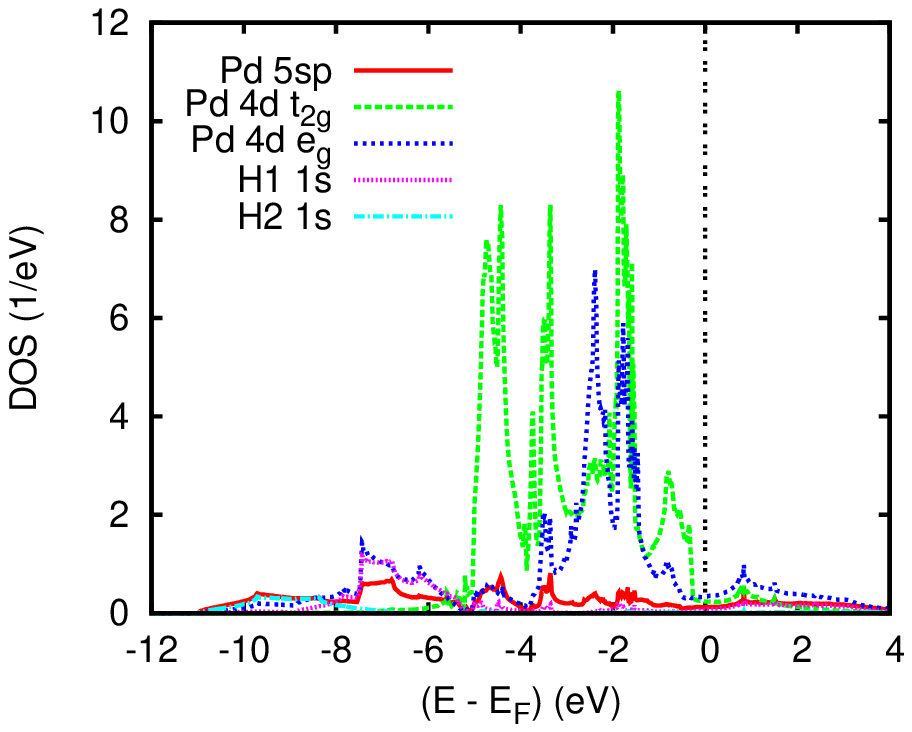}
\includegraphics[width=\columnwidth]{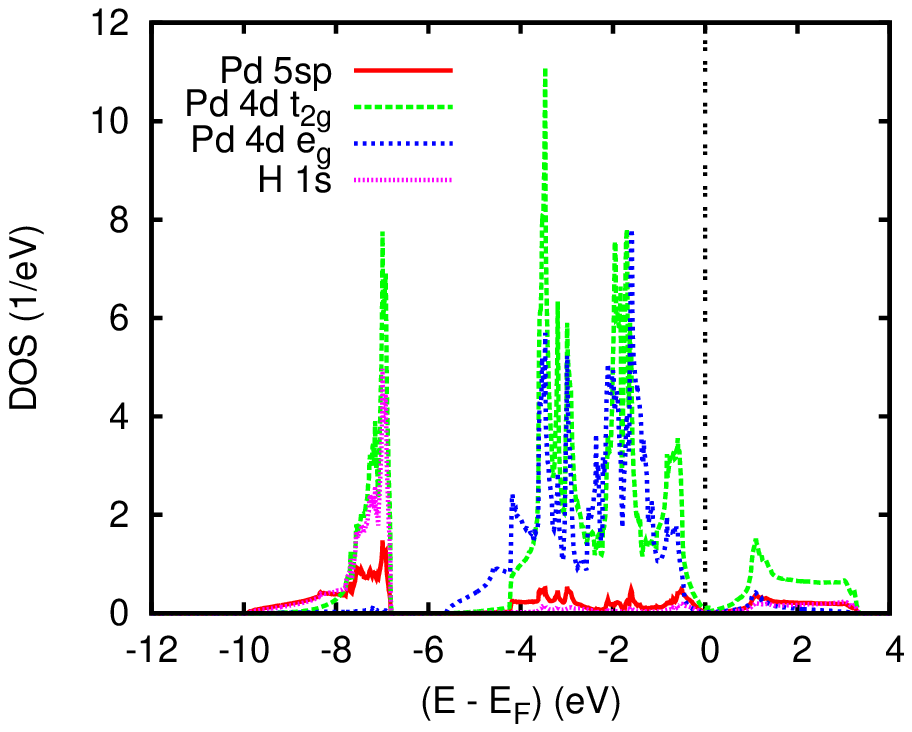}
\caption{Partial densities of states of $ {\rm Pd_3H_4} $ in the rocksalt 
         (H in octahedral site, top) and zincblende (H in tetrahedral 
         site, bottom) structure.}
\label{fig:dosPd3H4}
\end{figure}

The results of the subsequent calculations of the electronic structure 
using the ASW code are displayed in Fig.\ \ref{fig:dosPd3H4}. 
Here, from the distinction of the $ \sigma $- and $ \pi $-bonding manifolds 
of the Pd $ 4d $ states, the strong $ \sigma $-type overlap of the 
H $ 1s $ orbitals with the $ e_g $ and $ t_{2g} $ orbitals, respectively, 
in the rocksalt and zincblende structures becomes obvious. In addition, we 
point out that removal of the Pd atoms at the corners of the cube leads to 
two different types of hydrogen atoms in the rocksalt structure, namely, H1 
atoms located in the middle of the edges and H2 at the center of the cube. 
In contrast, in the zincblende-derived vacancy structure, still all hydrogen 
atoms are equivalent by symmetry. This has interesting consequences for the 
electronic states. While for the zincblende-derived structure only one 
split-off band in the energy range from $ -10 $\,eV to $ -7 $\,eV is 
observed, the split-off band in the rocksalt structure falls into two 
parts: The rather weak low-energy part between $ -11 $\,eV to $ -7.5 $\,eV 
arises from overlap of the Pd $ 4d $ states with the $ 1s $ states of the 
single central H2-type atoms, whereas the strong high-energy part between 
$ -7.5 $\,eV to $ -5 $\,eV is due to the overlap with the orbitals of the 
three H1-type atoms. This difference in $ d $-$ s $ splitting is easily 
explained from the fact that the H2-type atoms at the center of the cubic 
cell have six Pd neighbors, whereas the H1-type atoms at the edges have 
only four such neighboring Pd atoms. As a consequence, these latter atoms 
experience less overlap with the $ 4d $ orbitals. Finally, we mention the 
dip of all partial densities of states just above the Fermi energy, which 
for the dihydride was observed at the Fermi energy and attributed to the 
complete occupation of the $ d $ shell and the onset of mainly 
$ 1s $-derived bands. 

Finally, partial covalence energies were calculated and could be well 
understood from those obtained for the monohydrides and dihydride as 
shown in Figs.\ \ref{fig:coopPdPdH} and \ref{fig:coopPdH2} taking into 
account the increased H:Pd ratio as compared to PdH and the resulting 
appearance of the additional group of bands as discussed in Sec.\ 
\ref{sec:pdh2}. In particular, while strong Pd $ 4d $-H $ 1s $ bonding 
stabilizes the structure, antibonding $ d $-$ d $ interactions tend to 
undermine the overall stability.

\section{Summary}
\label{sec:summ}

In conclusion, three different types of palladium hydrides with varying 
hydrogen concentration have been investigated by first principles 
electronic structure calculations using two complementary tools, namely, 
the VASP and the ASW code. For the monohydride PdH the rocksalt structure 
has been identified as the ground state in agreement with previous 
experimental and theoretical work. However, were it not for the zero-point 
energy, the zincblende structure would be more stable. In particular, 
the strong difference in zero-point energy results from the larger 
voids available for the hydrogen atoms and consequently much softer 
Pd-H bonds in the rocksalt structure as compared to the zincblende 
and wurtzite structure.  As compared to Pd metal, insertion of hydrogen 
into the face-centered cubic lattice leads to the formation of bonding 
and antibonding $ 4d $-$ 1s $ states, which are found below and above 
the remaining Pd $ 4d $ bands, respectively. Especially, the bonding 
split-off bands are easily identified and underpinned by recent XPS and 
UPS data. In addition, hydrogenation leads to the upshift of the Fermi 
level to an energy range, where the density of states is low. As a 
consequence, the Pd-H bonding prevails over magnetovolume effects, leading 
to the suppression of magnetic order. For the dihydride $ {\rm PdH_2} $, 
the fluorite structure is found to be more stable than the pyrite structure 
albeit with a very small difference in heat of formation. The higher hydrogen 
content leads to an increased complexity of the band structure as compared 
to the metal and the monohydride due to the formation of a second split-off 
band below the central $ 4d $ bands and of an additional $ 1s $-derived band 
above this group. Finally, for supervacant $ {\rm Pd_3H_4} $, our 
calculations have established the rocksalt-type ground state. A detailed 
analysis of the electronic structures established the critical importance 
of strong covalent bonds of directional orbitals for the thermodynamic 
stability of all these hydrides.

{}

\end{document}